\documentclass{ifacconf}

\usepackage{graphicx}  
\usepackage{natbib}       
\usepackage{amsmath, amssymb, amsfonts} 
\usepackage{xcolor} 
\usepackage{comment} 
\usepackage{url}
\usepackage[newfloat]{minted}
\setminted{fontsize=\small, breaklines}
\definecolor{matlabblue}{rgb}{0, 0.4470, 0.7410}
\definecolor{matlabred}{rgb}{0.8500, 0.3250, 0.0980}
\definecolor{matlabgray}{rgb}{0.5, 0.5, 0.5}
\definecolor{matlablg}{rgb}{0.75, 0.75, 0.75}
\definecolor{matlabyellow}{rgb}{0.9290, 0.6940, 0.1250}
\definecolor{matlabpurple}{rgb}{0.4940, 0.1840, 0.5560}
\definecolor{matlab_br_red}{rgb}{1, 0, 0}
\definecolor{matlab_br_blue}{rgb}{0, 0, 1}
\definecolor{matlabblack}{rgb}{0, 0, 0}

\begin{document}
\begin{frontmatter}

\title{Identax: A Toolbox for Efficient Learning of Linear, Parameter-Varying, and Nonlinear Systems in JAX\thanksref{footnoteinfo}}

\thanks[footnoteinfo]{This work is funded by the European Union (ERC, COMPLETE, 101075836), by the Air Force Office of Scientific Research under award number FA8655-23-1-7061, and the MathWorks Inc. Views and opinions expressed are however those of the author(s) only and do not necessarily reflect those of the European Union or the European Research Council Executive Agency. Neither the European Union nor the granting authority can be held responsible for them.}
\author[First]{Roel Drenth}
\author[First]{Maarten Schoukens} 
\author[First,Second]{Roland T\'{o}th}

\address[First]{Control Systems Group, Dept. of Electrical Eng., Eindhoven University of Technology, Eindhoven, The Netherlands. (e-mail: \{r.drenth, m.schoukens, r.toth\}@tue.nl).}
\address[Second]{Systems and Control Laboratory, HUN-REN Institute for Computer Science and Control, Budapest, Hungary}
\begin{abstract} Accurate data-driven modelling of dynamical systems is increasingly important across engineering disciplines, yet state-of-the-art identification methods are often computationally demanding, opaque, or poorly scalable. We present Identax, a JAX toolbox for identifying linear, Linear Parameter-Varying (LPV) and nonlinear systems, designed as an extensible foundation for system identification research. Its modular design allows model structures, identification criteria and optimization methods to be extended independently. The first release provides state-space and Linear Fractional Representation (LFR) models, multiple-shooting criteria for parallel training, and a transparent optimization pipeline. On the Silverbox benchmark, an LPV-LFR model outperforms all previous black-box results while training in under three minutes on a CPU. On a control moment gyroscope, Identax matches state-of-the-art accuracy.
\end{abstract}
\end{frontmatter}

\section{Introduction}

It is of high importance to have accurate representations of dynamical systems across engineering domains. Depending on the application, models can be required for a wide range of methods, including model-based control design, simulation, prediction and forecasting or design optimization. No matter the application, the use of \emph{first-principles modelling} (FPM) is often too costly, or in the case of complex nonlinear systems, not sufficiently accurate. 

Instead, data-driven modelling, i.e., \emph{system identification} provides an attractive alternative, given sufficient data of the system-to-be-identified is available. The field of system identification has received much attention in research and encompasses many model classes. Well-developed methods exist for \emph{Linear Time-Invariant} (LTI) systems \citep{ljung1999,pintelon2012}, but methods for \emph{Linear Parameter-Varying} (LPV) \citep{toth2010} and more general \emph{nonlinear} (NL) classes \citep{schoukens2019} are still actively being developed.

Being inherently data-driven, system identification naturally relies on software implementations of its algorithms, leading to a diverse set of toolboxes and libraries. Classical approaches are supported by mature toolboxes such as the MATLAB System Identification Toolbox \citep{ljung2009}, SIPPY \citep{armenise2018} and for LPV models, LPVcore \citep{boef2021}. In recent years, motivated by advances in deep-learning, methods incorporating \emph{Neural Networks} (NN), such as SUBNET, implemented in deepSI \citep{beintema2023}, have been introduced and shown to be very effective. However, these methods are computationally demanding: training often takes tens of hours, which makes the usual iterative identification loop impractical.

In contrast, methods leveraging the JAX framework \citep{jax2018github}, which implements efficient automatic differentiation, just-in-time compilation and vectorization, have been shown to significantly reduce training times for frequency-domain techniques, tensor-networks and (nonlinear) state-space models in the freq-statespace \citep{floren2026freq}, tnkm \citep{saiapin2026} and jax-sysid \citep{bemporad2025a} libraries, respectively.

In this work, we are interested in efficient parametric modelling of linear, parameter-varying and nonlinear models using JAX. This has been previously implemented in jax-sysid, however with two important limitations: (i) its reliance on CPU-based optimization algorithms limits potential scalability of parallelization on GPU; (ii) lack of transparency of the optimization algorithm. As training and validation curves are not available, gaining insight into training behaviour is troublesome, complicating research.

With the goal of providing an alternative which solves these challenges, we present Identax\footnote{Available at \url{https://github.com/roel-drenth/identax}}, a modern toolbox for data-driven modelling of linear, parameter-varying and nonlinear systems, built with JAX. The goal of this toolbox is to provide a transparent, modular, scalable, and efficient solution for parametric modelling of dynamical systems. In its first release, Identax supports identification of \emph{State-Space} (SS) LTI models as well as LPV and NL models with SS and \emph{Linear Fractional Representation} (LFR), by \emph{Prediction Error Minimization} (PEM) \citep{ljung1999} methods. However, we emphasize that the toolbox has been designed with modularity and extensibility in mind, aiming to provide a foundation for future changes and extensions, e.g., alternative model structures, identification criteria, or model construction methods.

In the remainder of this paper, we introduce the current functionality of Identax and provide examples. First in Section \ref{sec: structures}, the current model structures are introduced, followed by explanation of the identification criteria in Section \ref{sec: criteria}. Next the optimization procedure is described in Section \ref{sec:construction}, followed by a description of additional features of the toolbox in Section \ref{sec: features}. Subsequently, benchmark problems are considered in Section \ref{sec:examples}. Finally, a future outlook of the toolbox is given in Section \ref{sec:outlook}, before conclusions are drawn in Section \ref{sec:conclusions}.

\section{Model Structures}\label{sec: structures}
We consider nonlinear \emph{discrete-time} (DT) data-generating systems with the following state-space representation:
\begin{equation}
\label{NLsys_eq}
\Sigma_\mathrm{o}: \left\{
\begin{aligned}
x_{k+1}  &= f(x_k,u_k),\\
y_k &= g(x_k,u_k) + v_k,
\end{aligned}
\right.
\end{equation}

where $k \in \mathbb{Z}$ is the discrete time, $x_k \in  \mathbb{X}\subseteq \mathbb{R}^{n_\mathrm{x}}$ is the state of the plant,  $u_k \in \mathbb{U} \subseteq \mathbb{R}^{n_\mathrm{u}}$ is the input signal, $y_k \in \mathbb{Y} \subseteq \mathbb{R}^{n_\mathrm{y}}$ is the system output, with output-additive white noise  $v_k \in \mathbb{V} \subseteq \mathbb{R}^{n_\mathrm{y}}$. The input $u_k$ is assumed to be quasi-stationary. The state transition function $f: \mathbb{X} \times \mathbb{U} \!\to\! \mathbb{X}$ and output function $g: \mathbb{X} \times \mathbb{U} \!\to\! \mathbb{Y}$ are considered to be deterministic and possibly nonlinear.

All considered models are DT and have output-error noise setting, with the following predictor: 
\begin{equation}
    \hat{y}_k = G_\theta u_k,
\end{equation}
where $\hat{y} \in \mathbb{R}^{n_\mathrm{y}}$ is the simulated output. As first model type, LTI-SS models with the following model equations are considered: 
\begin{equation}
\label{LTI_eq}
G_\theta: \left\{
\begin{aligned}
\hat{x}_{k+1}  &= A_x \hat{x}_k + B_u u_k,\\
\hat{y}_k &= C_y\hat{x}_k + D_{yu}u_k,
\end{aligned}
\right.
\end{equation}
for which the model parameters are collected in $\theta_G = \mathrm{vec}\left([A_x, B_u, C_y, D_{yu}]\right)$.

In Identax, models are implemented using Equinox \citep{kidger2021}. Each model class has an associated configuration dataclass. An example using LTI-SS models is given below: 
\begin{listing}[h]
\vspace{-6pt}
\begin{minted}{python}
key = jax.random.PRNGKey(42) # seed for init.
lti_cfg = idx.LinStateSpaceConfig(nx=nx, nu=nu, ny=ny)
lti_mod = idx.ModelLinStateSpace(lti_cfg, key)
\end{minted}
\label{lst:lti_model}
\vspace{-6pt}
\end{listing}\\
To represent LPV models, Identax implements LPV models as LFRs following the parameterizations proposed in \cite{drenth2026}. As an example, we consider LPV-LFR models with affine scheduling-dependency, described by:
\begin{equation}
\label{LPVLFR_plant}
G_\theta: \left\{
\begin{aligned}
\begin{bmatrix}
\hat{x}_{k+1} \\
\hat{z}_k\\
\hat{y}_k
\end{bmatrix} &=
\begin{bmatrix}
A_x & B_w & B_u\\
C_z & 0 & D_{zu}\\
C_y & D_{yw} & D_{yu}
\end{bmatrix}
\begin{bmatrix}
\hat{x}_k\\
\hat{w}_k\\
u_k
\end{bmatrix},\\
\hat{w}_k &= \Delta(\hat{p}_k)\hat{z}_k,
\end{aligned}
\right.
\end{equation}
with $D_{zw} = 0$, where $\Delta(\hat{p}_k)$ is diagonal with $r$ repeating values of $\hat{p}_k,$ i.e.,  
\begin{equation}
\label{diag_delta}
\Delta(\hat{p}_k) = \mathrm{diag} ( \hat{p}_{k,1} I_{r_1}, \dots, \hat{p}_{k,n_p} I_{r_{n_p}}),
\end{equation} and  $\hat{w}_k \in \mathbb{R}^{n_\mathrm{w}}$ and $\hat{z}_k \in \mathbb{R}^{n_\mathrm{z}}$ represent the latent variables of the LFR. The scheduling signal is generated by a scheduling map: $\hat{p}_k = \hat{\psi}_\theta(\hat{x}_k,u_k, d_k)$, in which $d_k$ represents exogenous signals, used for externally-scheduled LPV models. Identax currently features scheduling maps parameterized as \emph{Feedforward Neural Networks} (FNN), possibly with linear bypass (ResNets). The parameters of \eqref{LPVLFR_plant} are collected in  $\theta_G = \mathrm{vec}([A_x, B_w, B_u, C_z, D_{zu}, C_y,\allowbreak D_{yw},\allowbreak D_{yu}])$, and the parameters of the scheduling map in $\theta_\psi$.

Again, models are constructed using a configuration. LPV models have a separate configuration for their scheduling maps. LPV-LFR models can be instantiated as follows:
\begin{listing}[h]
\vspace{-6pt}
\begin{minted}{python}
sched_cfg = idx.SchedulingMapConfig(
    network="ResNet",
    layers=2,
    neurons_per_layer=6)
lpv_cfg = idx.AffineLPVLFRConfig(
    nx=nx, nu=nu, ny=ny,
    npar=npar,         # number of sched. signals
    nreps=nreps,       # number of repetitions in Delta
    nd=0,              # no exo. scheduling
    SchedulingMap=sched_cfg,
lpv_mod = idx.ModelAffineLPVLFR(lpv_cfg, key)
\end{minted}

\label{lst:lpv_model}
\vspace{-6pt}
\end{listing}\\
Finally, the class of NL-LFR models is considered, represented by the following model equations:
\begin{equation}
\label{NLLFR_eq}
G_\theta: \left\{
\begin{aligned}
\begin{bmatrix}
\hat{x}_{k+1} \\
\hat{z}_k\\
\hat{y}_k
\end{bmatrix} &=
\begin{bmatrix}
A_x & B_w & B_u\\
C_z & D_{zw} & D_{zu}\\
C_y & D_{yw} & D_{yu}
\end{bmatrix}
\begin{bmatrix}
\hat{x}_k\\
\hat{w}_k\\
u_k
\end{bmatrix},\\
\hat{w}_k &= \hat{\eta}(\hat{z}_k),
\end{aligned}
\right.
\end{equation}
with $D_{zw} = 0$ to ensure well-posedness. The feedback nonlinearity $\hat{\eta}$ is parameterized as FNN or ResNet. The trainable parameters of the NL-LFR are collected in $\theta_G = \mathrm{vec}\left([A_x, B_w, B_u, C_z, D_{zu}, C_y, D_{yw}, D_{yu}]\right)$ and the parameters of the feedback nonlinearity in $\theta_\eta$.

Identax supports jointly estimating a \emph{subspace encoder} \citep{beintema2023} alongside nonlinear models. These encoders can be used for estimation of initial states in \emph{Encoder-Based Multiple-Shooting} (EBMS), further detailed in Section \ref{sec: criteria}. Subspace encoders map previous samples to an initial state estimate:
\begin{equation}\label{eq: encoder}
    \hat{x}_{0,t} =  \phi(y_{t-n_\mathrm{hist}:t-1}, u_{t-n_\mathrm{hist}:t-1}, d_{t-n_\mathrm{hist}:t-1}).
\end{equation}
Identax implements encoder networks parameterized as FNN or ResNet as an optional trainable network for nonlinear and LPV models. Its parameters are collected in the parameter vector $\theta_\phi$.

Encoders are initialized as part of a model using an \mintinline{python}|idx.StateEncoderConfig()|. For example, an NL-LFR with encoder network can be created as follows:\\
\begin{listing}[!ht]
\vspace{-6pt}
\begin{minted}{python}
nllfr_cfg = idx.NLLFRConfig(
    nx=nx, nu=nu, ny=ny, nw=1, nz=1,
    DeltaMap=idx.NetworkConfig(neurons_per_layer=6),
    StateEncoder=idx.StateEncoderConfig(n_hist = nx*2+1))
nllfr_mod = idx.ModelNLLFR(nllfr_cfg, key)
\end{minted}
\label{lst:nl_model}
\vspace{-6pt}
\end{listing}\\
where \mintinline{python}|DeltaMap()| represents the feedback nonlinearity of the NL-LFR.

Besides individual models Identax supports ensembles of models, i.e., a group of models sharing the same structure but with individual parameters. Currently, this can be used to train a group of models with different initializations. Ensemble models are created using the same configuration objects as their non-ensemble counterparts, e.g.,:
\begin{listing}[!ht]
\vspace{-6pt}
\begin{minted}{python}
nllfr_ens = idx.EnsembleNLLFR(n_models=10, config=nllfr_cfg, key=key)
\end{minted}
\label{lst:ensemble_model}
\vspace{-6pt}
\end{listing}\\
creates an ensemble of 10 NL-LFR models.
\section{Identification Criteria} \label{sec: criteria}
Currently, Identax implements a PEM cost function. For the model structures described in Section \ref{sec: structures}, this is equivalent to the simulation error cost. Additionally, Identax supports identification based on multiple datasets or \emph{experiments}, with the following associated cost function:
\begin{equation}\label{eq:SE-cost}
V^\mathrm{MSE}(\theta) = \frac{1}{N_\mathrm{tot}}
  \sum_{e=1}^{E} \sum_{k=0}^{N_e-1} \big\|y_{e,k}-\hat{y}_{e,k}\big\|_2^2,
\end{equation}
where $N_\mathrm{tot} = \sum_{e=1}^E N_e$ represents the sum of samples $N_e$ across all $E$ experiments, $y_{e,k}$ denotes sample $k$ of experiment $e$. The cost function \eqref{eq:SE-cost} can be extended with a regularization term. Identax currently supports $\ell_2$-regularization of the model parameters:
\begin{equation}
    V^\mathrm{reg}(\theta) = \rho\|\theta\|_2^2
\end{equation}
where $\rho$ is a user-chosen regularization weight and $\theta$ is a vector collecting all trainable parameters of a model.

To further improve conditioning of the cost, and enable parallelism, \emph{Multiple Shooting} methods \citep{bock1983} are implemented in Identax. Currently, \emph{Unconstrained Multiple Shooting} (UMS) \citep{decuyper2020} and EBMS \citep{beintema2023} are supported. In multiple-shooting, the cost \eqref{eq:SE-cost} is replaced by the \emph{truncated prediction error} in \eqref{eq:MS}. For clarity, the single-experiment cost is shown, however multi-experiment multiple-shooting is supported in Identax.
\begin{equation}\label{eq:MS}
 V^\mathrm{MS}(\theta) = \frac{1}{T R}
  \sum_{t\in\mathcal I}\sum_{k=0}^{T-1}\big\|y_{t+k}-\hat y_{t+k|t}\big\|_2^2,
\end{equation}
with $T$ representing the truncation or sub-record length, $R$ the number of sub-records and $\mathcal{I}$ the set of indices of each sub-record. Currently, Identax implements splitting records into sub-records with equidistant spacing, and user-chosen overlap of sub-records $n_o$, i.e.,
\begin{equation}
    \mathcal{I} = \left\{n_\mathrm{hist} + j\,(T - n_\mathrm{o}) \;\middle|\; j = 0, 1, \dots, R - 1 \right\},
\end{equation}
where $n_\mathrm{hist} $ in \eqref{eq: encoder} is the number of past samples used for initial state estimation in EBMS.

As Identax supports parallelism on CPU and GPU, the use of multiple-shooting methods can reduce the wall-clock time of training substantially, compared to single shooting which requires recurrently evaluating the model equations for the full record length, resulting in wall-clock computation time $\mathcal{O}(N)$, omitting additional overhead this can theoretically be reduced to approximately $\mathcal{O}(T)$, as long as the number of parallel devices is at least $R$. For large $R$, this makes training amenable to the use of GPU devices. A deeper comparison of UMS and EBMS for LPV-LFR models is performed in \cite{drenth2026b}

\section{Model Construction}\label{sec:construction}
To construct models, given the criteria in Section \ref{sec: criteria}, Identax currently implements a gradient-based optimization procedure. A two-step procedure, based on the method proposed in \cite{bemporad2025a}, is implemented. In the first step of this procedure, a fixed number of iterations is performed using the Adam optimizer \citep{kingma2017}, providing a warm-start for the second step in the procedure. In this step, the quasi-Newton \emph{limited memory Broyden-Fletcher-Goldfarb-Shanno} (L-BFGS) algorithm \citep{liu1989} is used to converge to a final minimum of the cost.

The optimization procedure is implemented using the Optax library \citep{deepmind2020jax}, which provides JAX implementations of a wide-range of optimizers, including Adam and L-BFGS. The required gradient calculations are performed using the built-in automatic differentiation tools of JAX. In addition to the off-the-shelf optimizers of Optax, Identax provides several extra features. For L-BFGS, it implements convergence-based stopping conditions on the gradient norm and on the change in loss between iterations. Identax also supports early stopping based on hold-out validation, which returns the model from the optimization step with the lowest validation cost.

\subsection{Learning model parameters}

Identax aims to give full control while providing sensible defaults. To achieve this, the optimization procedure is customized using configuration objects which expose most options but provide robust default settings. An example of how optimization can be configured is given below: 
\begin{listing}[h]
\vspace{-6pt}
\begin{minted}{python}
optim_cfg = idx.OptimConfig(
    adam_epochs=1000, lbfgs_epochs=2000,
    shooting_type="encoder", # use EBMS
    T=1000, overlap=100,     # length and overlap for EBMS
    l2_rho=1e-6,             # l2 regularization coeff.
    adam_lr=1e-4,            # adam learning rate
    lbfgs_max_ls=10,)        # max linesearches 
\end{minted}
\label{lst:optimconfig}
\vspace{-6pt}
\end{listing}

Note that arguments not supplied to \mintinline{python}|idx.OptimConfig()| use pre-defined defaults. Having constructed a config, a model can be trained as follows:
\begin{listing}[h]
\vspace{-6pt}
\begin{minted}{python}
new_mod, x0, history = idx.optimize(
    mod, data_train, x0_init, optim_cfg,
    val_data=val_data) # optional for early stopping
\end{minted}
\label{lst:optim_func}
\vspace{-6pt}
\end{listing}\\
Here, \mintinline{python}|idx.optimize()| returns the trained model, its learned initial states and a history object. The history object contains information such as training time, stopping conditions and loss curves, including validation loss if \mintinline{python}|val_data| was supplied.

\subsection{Learning initial states}
Once \mintinline{python}|idx.optimize()| has learned a model it can be used to estimate the initial states on other datasets. In Identax, this is done by minimizing \eqref{eq:SE-cost} with respect to $x_0$, using L-BFGS. This is implemented using \mintinline{python}|idx.learn_initial_state()| and can be called as follows:
\begin{listing}[h]
\vspace{-6pt}
\begin{minted}{python}
x0_val, history = idx.learn_initial_state(mod, val_data, x0_init, lbfgs_epochs=100, T=50)
\end{minted}
\label{lst:learn_init}
\vspace{-6pt}
\end{listing}\\
where $T$ truncates the dataset to only use the first $T$ values in the optimization criterion.

\section{Additional Features}\label{sec: features}
There are a number of features implemented in Identax to improve usability of its methods. First, we consider data-handling. Instead of manually handling input, output and possibly scheduling data, experiment data is captured in \mintinline{python}|idx.iddata()| objects, which can subsequently be used for training, validation, or testing. These objects are easily created:
\begin{listing}[h]
\vspace{-6pt}
\begin{minted}{python}
train_data = idx.iddata(U_train, Y_train, D=D_train)
\end{minted}
\label{lst: iddata}
\vspace{-6pt}
\end{listing}\\
where \texttt{U\_train}, \texttt{Y\_train} and \texttt{D\_train} have shapes $N \times n_u, \ N \times n_y$ and $N \times n_d$. To handle multiple experiments, e.g., in multi-experiment training, the \mintinline{python}|idx.iddata_collection()| object exists, created by supplying a list of \texttt{iddata} objects:
\begin{listing}[h]
\vspace{-6pt}
\begin{minted}{python}
exp1 = idx.iddata(U=U1, Y=Y1)
exp2 = idx.iddata(U=U2, Y=Y2)
exp3 = idx.iddata(U=U3, Y=Y3)
multi_exp = idx.iddata_collection([exp1, exp2, exp3])
\end{minted}
\label{lst: iddata_collection}
\vspace{-6pt}
\end{listing}\\
After creation, the \texttt{iddata\_collection} object can directly be supplied to e.g., optimize() to train a model.

To improve conditioning of optimization algorithms, it is common practice to normalize datasets. In Identax, this is handled by \mintinline{python}|idx.DataScaler()| objects. Each model has an associated scaler, which collects the type of transformation per signal (input, output, and scheduling). For example, a standard scaler with mean and standard deviation can be estimated from data, and used to transform data as follows:
\begin{listing}[h]
\vspace{-6pt}
\begin{minted}{python}
cfg = idx.DataScalerConfig(type_u='standard', type_y='standard')
scaler = idx.DataScaler.from_data(train_data, cfg)
train_scaled = scaler.transform(train_data)
\end{minted}
\label{lst: scaling}
\vspace{-6pt}
\end{listing}\\
By default, Identax initializes models with standard scalers, and estimates mean and standard deviation from training data when \texttt{optimize()} is first called. Users are not required to scale data manually, user-facing functions handle unscaled data, transform it using the \texttt{DataScaler()} of a model, and handle scaled data internally for e.g., simulation and optimization.

To enable users to save and load models, serialization of models is implemented. Models can be saved and loaded as follows:
\begin{listing}[h]
\vspace{-6pt}
\begin{minted}{python}
save_model("saved_models/lti_model.zip", model, overwrite=True)
loaded = load_model("saved_models/lti_model.zip", ModelLinStateSpace)
\end{minted}
\label{lst: serialize}
\vspace{-6pt}
\end{listing}\\
The save file contains a \texttt{config.json} file with the configuration of the model and a \texttt{weights.eqx} file saving all the model parameters, loading a model requires specifying the model class.

\section{Examples}\label{sec:examples}
\subsection{Silverbox}
As a first example we consider the Silverbox \citep{wigren2013}. The Silverbox is an electronic implementation analogous to a nonlinear mass-spring-damper with stiffening spring, also known as a forced Duffing oscillator. It is a simple nonlinear system commonly used as benchmark in nonlinear system identification \citep{schoukens2026}, as it can be represented as a second-order model with cubic nonlinearity.

The data of the Silverbox consists of two sections. One section is excited by a filtered white-noise input with increasing amplitude, called the "arrow". The second part consists of  10 realizations of a random-phase odd  multisine, exciting frequencies up to 200 Hz. The official data split of \cite{schoukens2026} splits the multisine segment into  65062 samples for training and validation and 21688 samples for testing. We further perform an 80/20 training/validation split, resulting in training and validation records of respectively 52049 and 13013 samples. The arrow segment is solely used for testing.

We train each of the model types described in Section \ref{sec: structures}, starting with LTI-SS models. We consider models with $n_x = 2$, matching the state of the nonlinear data-generating system. For the LPV-LFR model $n_p = 1$ and $r = 1$. $\hat{\psi}_\theta$ is parameterized as ResNet with 2 hidden layers and 6 nodes per layer. For the NL-LFR $n_w =1$ and  $n_z = 1$. The feedback nonlinearity $\hat{\eta}_\theta$ follows  the same parameterization as $\hat{\psi}$.

First, LTI-SS models are trained for 1000 iterations of Adam and L-BFGS each. Subsequently, LPV-LFR and NL-LFR models are initialized following the initialization procedures of \cite{drenth2026b} and \cite{schoukens2020} respectively. To illustrate functionality of multiple shooting, LPV-LFR and NL-LFR models are trained once on the full trajectory, and once using UMS, with truncation length $T=200$ and no overlap. All LFR models are trained for 1000 Adam iterations followed by at most 20000 iterations of L-BFGS. A regularization weight of $\rho = 10^{-6}$ is used. The validation dataset is used for early stopping. The first 50 samples of the validation and test dataset are used for estimating the initial state using \mintinline{python}|idx.learn_initial_state()|, these samples are excluded for evaluating error. All models are trained on an Intel Core I7-14700K CPU.

To evaluate the models, the \emph{Root-Mean-Squared Error} (RMSE) is used, defined as $\mathrm{\epsilon}_{\mathrm{RMS}} = \sqrt{1/N}\left\| y - \hat{y} \right\|_2$.

The results, shown in Table \ref{tb: silverbox_results}, show that both LFR models are capable of obtaining accurate results on the Silverbox benchmark, in fact, the results on the arrow test dataset show that the LPV-LFR model trained with UMS outperforms all previous benchmark results using strictly black-box methods such as TNKM \citep{saiapin2026} and SUBNET \citep{beintema2021}. Only the \emph{Polynomial Nonlinear State-Space} (PNLSS) \citep{paduart2010} method, which leverages prior knowledge of the nonlinearity, obtains a better result. Models trained with UMS show slightly lower training times than their single-shooting counterparts, larger reduction in training time is expected as the number of records becomes even larger and a GPU is used. Nonetheless, all models are trained in less than three minutes, illustrating the computational efficiency of the implementation.
\begin{table}[!t]
\begin{center}
\caption{Results on the Silverbox benchmark, shown in $10^3\times$ RMSE. Litereature values taken from \cite{paduart2010,saiapin2026,beintema2021}.}\label{tb: silverbox_results}
\begin{tabular*}{\columnwidth}{@{\extracolsep{\fill}} l c c c c r}
Model  & train & test (multi.) & test (arrow) & time \\\hline

LTI-SS & 6.90 & 6.96 & 14.36 &   3 s\\
LPV-LFR & $\mathbf{0.21}$ &\textbf{0.22}& 0.38 &   179 s\\
LPV-LFR (UMS) & 0.22 &0.22& 0.29 &   170 s\\
NL-LFR   & 0.24 &0.25& 0.42  & 136 s \\
NL-LFR (UMS) & 0.24 & 0.25 & 0.54  & 126 s \\
PNLSS & - & -& \textbf{0.26}  & -  \\
TNKM & - & 0.40& 0.88  & $\leq 60$ s\footnotemark \\
SUBNET &  - & 0.36 & 1.6  & $\leq 96$ h\footnotemark[\value{footnote}] \\\hline
\end{tabular*}
\end{center}
\end{table}
\footnotetext{Trained on different hardware, only used as indication.}
\subsection{Control Moment Gyroscope}
\begin{figure}[!t]
\begin{center}
\includegraphics[width=2.4cm]{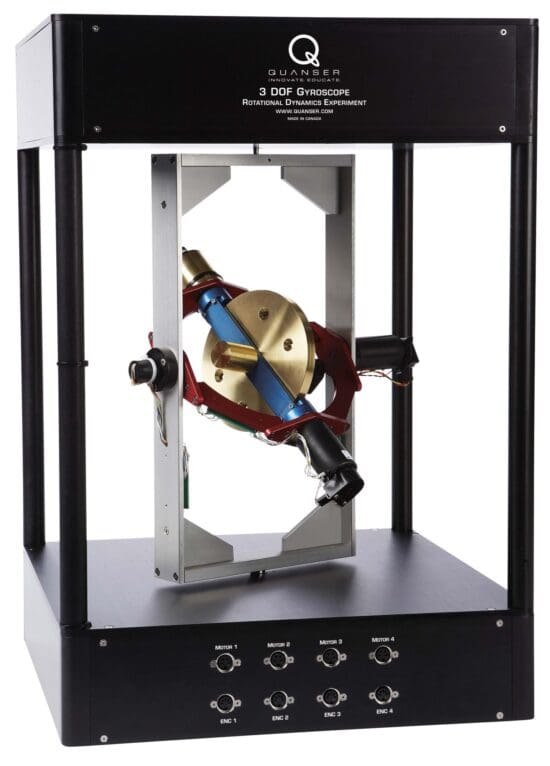}    
\caption{Control moment gyroscope} 
\label{fig: gyroscope}
\end{center}
\end{figure}

As second example, the \emph{Control Moment Gyroscope} (CMG) introduced in \cite{verhoek2022} is considered. The CMG consists of an inner disk mounted inside a series of gimbals of blue, red, and silver colors, shown in Figure \ref{fig: gyroscope}. The disk and gimbals are actuated by DC motors, and the angular positions are measured by encoders, denoted as $q_1, q_2, q_3$ and $q_4$ for the disk, blue, red, and silver gimbals respectively. The dataset is generated by a high-fidelity simulator, in which the red gimbal is locked, i.e., its angle is fixed at $q_3 =0$. The motor current of the blue gimbal, $i_2$, is considered as input and the angular velocity of the silver gimbal, $\dot{q}_4$, is considered as output. The angular velocity of the disk, $\dot{q}_1$ is controlled independently and follows a changing reference velocity. For small angles of $q_2$ the resulting dynamics can be considered as a parameter-varying SISO system from 
$i_2$ to $\dot{q}_4$ with a scheduling signal $\dot{q}_1$, or as a MISO system with $i_2$ and $\dot{q}_1$ as inputs and $\dot{q}_4$ as output.

The data of the CMG consists of a training and a test dataset, of 10000 and 30000 samples respectively. The input signal $i_2$ consists of sinusoidal signals with additive noise. The reference trajectories of $\dot{q}_1$ are multi-level reference signals between 30 and 40 rad/s. The resulting output data has artificial Gaussian white-noise added resulting in a signal-to-noise ratio of 35 dB. The data is sampled at 100 Hz.

First, we show how LPV-LFR models are capable of handling external scheduling signals, by considering a SISO LPV-LFR model with external scheduling signal $d = \dot{q}_1$. To initialize the LPV-LFR model an LTI model with $n_u = 1, n_y = 1$ and $n_x =5$ is trained on IO data for 1000 Adam and L-BFGS iterations. Subsequently, this model is used to initialize an LPV-LFR model, according to the initialization procedure in \cite{drenth2026b}, with $n_\mathrm{p} =3$ and $r=1$. This model is trained for 1000 iterations with Adam and at most 20000 with L-BFGS.

Subsequently, to train an NL-LFR model, the MISO case, where $i_2$ and $\dot{q}_1$ are inputs, is considered. Again, an LTI-SS model is trained for 1000 Adam and L-BFGS iterations, after which the NL-LFR model with $n_x = 5, n_w = 3$ and $n_z=3$ is initialized and trained for 1000 Adam and at most 20000 L-BFGS iterations.

To evaluate the accuracy of the models, we compare our results with those of \cite{bemporad2025b}, in which accuracy is assessed using the \emph{Best Fit Rate} (BFR): 
\begin{equation}\label{eq: BFR}
   \epsilon_\mathrm{BFR} = \max\left\{1- \frac{\sum^{N-1}_{k=0}\|y_k - \hat{y}_k\|_2^2}{\sum^{N-1}_{k=0}\|y_k - \bar{y}\|_2^2},\  0 \right\}\cdot 100\%.
\end{equation}
The results are summarized in Table \ref{tb: cmg_results}, which shows that the methods in Identax are on par with methods implemented in jax-sysid in terms of training time and achievable accuracy. LPV-SUBNET achieves slightly better results than LPV-LFR, at the cost of a much longer training time. However it should be noted that LPV-LFR and NL-LFR methods are trained without access to test data, unlike LPV-SS and NL-SS methods of \cite{bemporad2025b}, which use 100 and 500 different initializations respectively, selecting the best result on test data. Our method only uses training data on a single initialization. Furthermore, LPV-SUBNET uses early stopping based on the test data, which is not used for training LPV-LFR and NL-LFR models using Identax.

\begin{table}[!b]
\begin{center}
\caption{Results on the CMG benchmark, shown in BFR.}\label{tb: cmg_results}
\begin{tabular*}{\columnwidth}{@{\extracolsep{\fill}} l c c c r}
Model  & train & test &  time \\\hline
LTI-SS $(n_u = 1)$ & 35.51 & 25.73 &    2 s\\
LTI-SS $(n_u = 2)$ & 36.35 & 25.13 &    2 s\\
LPV-LFR & \textbf{97.63} & 96.22 &    51 s\\
NL-LFR   & 97.03 & 95.76 & 46 s \\
LPV-SS \citep{bemporad2025b}   & 97.61 & $\mathbf{96.50}$ & 48 s\footnotemark[\value{footnote}] \\
NL-SS \citep{bemporad2025b}  & 96.75 & 96.12 & 42 s\footnotemark[\value{footnote}] \\
LPV-SUBNET \citep{verhoek2022} & 97.27 & 96.40 & $\approx$ 10 h\footnotemark[\value{footnote}] \\ \hline
\end{tabular*}
\end{center}
\end{table}

\begin{figure}[!b]
\begin{center}
\includegraphics[width=\columnwidth]{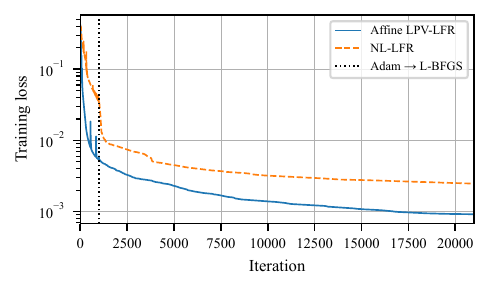}    
\caption{Training losses of LPV-LFR and NL-LFR models on CMG.} 
\label{fig: cmg_train_loss}
\end{center}
\end{figure}

Finally, to illustrate the transparency of the optimization process, the training losses of LPV-LFR and NL-LFR models on the CMG are displayed in Figure \ref{fig: cmg_train_loss}. It should be noted that jax-sysid does not expose the loss history of its L-BFGS stage. The training losses show a sharp change in convergence rate as the optimization switches from Adam to L-BFGS.

\section{Outlook}\label{sec:outlook}
In the future, we aim to extend Identax beyond the baseline functionality described in this work. First, we aim to introduce model structures capable of embedding prior knowledge through \emph{Model Augmentation}, such as the LFR-based model augmentation structure proposed in \cite{hoekstra2025}. Moreover, current model structures are limited to representing noise models with output-error structure. These will be extended to more complex NL noise representations as presented in \cite{bemporad2025b}. Furthermore, we aim to extend regularization flexibility by adding weighted $\ell_2$-regularization. To aid users in model-order selection, group-lasso regularization will be added. Finally, to bridge the gap from modelling to control of LPV models, we aim to provide an interface to LPVcore \citep{boef2021}, which implements modern analysis and synthesis tools for LPV models.

\section{Conclusion}\label{sec:conclusions}
This work has introduced Identax, a toolbox for efficient learning of linear, parameter-varying and nonlinear systems with JAX. Through its scalability, extensibility, transparency and efficiency, the toolbox can support research and application of state-of-the-art identification methods. The current functionality of the toolbox has been described, and the efficient modelling capability has been shown on two benchmarks, outperforming existing state-of-the-art black-box modelling methods on the Silverbox in terms of achieved accuracy. The initial release of the toolbox provides a solid foundation in terms of functionality, but will be extended in the future.

\section*{Declaration of generative AI and AI-assisted technologies in the writing process} During the preparation of this work the authors used \emph{Claude} (Anthropic) to improve readability and language. After using this tool, the authors reviewed and edited the content as needed and take full responsibility for the content of the published article.
\bibliography{ifacconf} 
\end{document}